\documentclass[showpacs,pra,aps,superscriptaddress,twocolumn,floatfix]{revtex4}

\usepackage{amsmath}
\usepackage{amssymb,mathrsfs}
\usepackage{graphicx}

\newcommand {\prt}{\partial}

\begin{document}

\title{Nonlinear waves of polarization in two-component Bose-Einstein
  condensates}

\author{A.~M.~Kamchatnov}
\affiliation{Institute of Spectroscopy,
  Russian Academy of Sciences, Troitsk, Moscow, 142190, Russia}
\author{Y.~V.~Kartashov}
\affiliation{Institute of Spectroscopy,
  Russian Academy of Sciences, Troitsk, Moscow, 142190, Russia}
\author{P.-\'E. Larr\'e}
\affiliation{Univ. Paris Sud, CNRS, Laboratoire de Physique Th\'eorique
et Mod\`eles Statistiques, UMR8626, F-91405 Orsay, France}
\author{N. Pavloff}
\affiliation{Univ. Paris Sud, CNRS, Laboratoire de Physique Th\'eorique
et Mod\`eles Statistiques, UMR8626, F-91405 Orsay, France}

\begin{abstract}
  Waves with different symmetries exist in two-component Bose-Einstein
  condensates (BECs) whose dynamics is described by a system of
  coupled Gross-Pitaevskii (GP) equations. A first type of waves
  corresponds to excitations for which the motion of both components
  is locally in phase. In the second type of waves the two components
  have a counter-phase local motion.  In the case of different values
  of inter- and intra-component interaction constants, the long
  wave-length behavior of these two modes corresponds to two types of
  sound with different velocities.  In the limit of weak nonlinearity
  and small dispersion the first mode is described by the well-known
  Korteweg-de Vries (KdV) equation. We show that in the same limit the
  second mode can be described by the Gardner (modified KdV) equation,
  if the intra-component interaction constants have close enough
  values. This leads to a rich phenomenology of nonlinear excitations
  (solitons, kinks, algebraic solitons, breathers) which does not
  exist in the KdV description.
\end{abstract}


\maketitle

\section{Introduction}

Two-component Gross-Pitaevskii (GP) equations describe the evolution
of nonlinear excitations in various physical systems.  Apparently, it
first appeared in nonlinear optics as a ``vector nonlinear
Schr\"odinger equation'' describing self-interaction of
electromagnetic waves with account of their polarization (see, e.g.,
\cite{manakov,sp-1997,ka-2003}). In the case of equal nonlinearity
constants, this system is completely integrable by the inverse
scattering transform method \cite{manakov} and many particular
solutions have been found (see, e.g., \cite{ps-2000} and references
therein). Taking birefringence effects into account \cite{menyuk-1987}
leads to even richer dynamics, as confirmed by experiments on
propagation of light pulses in fibers \cite{fiber-1999}.  The recent
surge of interest in vector solitons was caused by the realization of
spinor atomic BECs (see the reviews \cite{ku-2012,Sta13} and
references therein) as well as micro-cavity polariton
condensates \cite{hivet-2012}.

From the physical point of view, a specific feature of two-component
condensates is the existence of two types of elementary excitations.
In one mode both condensates move locally in phase. In the case of a
small amplitude potential flow this corresponds to usual sound waves
consisting in density oscillations. In another ``polarization'' mode
the two components move in counter-phase in such a way that, in some
situations, the total density remains constant in spite of the
excitation of relative motion of the condensate
components. Correspondingly, these two modes can have different values
of the sound velocities and, due to different symmetries, their
excitation demand different methods \cite{Fla12,lpk-2013}.  In some
sense, this is analogous to the situation observed for the first and
second sound in superfluid HeII: the second sound which corresponds to
a temperature (and entropy) wave cannot be excited by oscillations of
the container wall, contrarily to the usual density waves associated
with the first sound; see, e.g., \cite{ll-6}.

In the present paper, we study the weakly nonlinear evolution of these
two modes in presence of small dispersive effects. Whereas in this
limit the density waves are described by the standard KdV equation
which accounts for quadratic nonlinearities in the wave amplitude, the
polarization mode is much more peculiar. We show that there are
situations where one has to take into account third order
nonlinearities in the wave amplitude and that the corresponding
nonlinear polarization wave is then described by the Gardner
equation. As is well known, this is a quite generic equation which arises
when the coefficient in the quadratic nonlinearity term is small and
when the wave amplitude has the same order of magnitude as this
coefficient. In particular, the Gardner equation (and also the
modified KdV equation which shares strong similarities with it) have
been derived for interface waves in stratified fluid dynamics
\cite{ky-1978,mh-1987}, lattice dynamics described by the discrete
nonlinear Schr\"odinger equation \cite{ksk-2004}, and quantum dynamics
of condensates in optical lattices \cite{DemMal2011}. The Gardner
equation has a wide spectrum of different nonlinear excitations
\cite{kamch-2012} which can be generated by the flow of fluid past an
obstacle \cite{kamch-2013}. We expect that the phenomenology
associated to this rich dynamics can be observed in experiments
with flows of BECs.

\section{Main equations and linear waves}

We consider a two-component condensate confined in a one-dimensional
structure (e.g., a cigar-shaped trap for atomic condensates or a
quantum wire embedded into an elongated cavity for polariton
condensate case) in which the BEC is described by a one-dimensional
(1D) two-component order parameter $(\psi_+(x, t),\psi_-(x, t))$ whose
dynamics is modeled by a set of coupled Gross-Pitaevskii equations
\begin{equation}\label{eq1}
i \partial_t \psi_{\pm} + \tfrac12 \partial_{xx}^2\psi_{\pm}
- \Big[(\alpha_1\pm\delta)|\psi_\pm|^2+\alpha_2|\psi_{\mp}|^2\Big] \psi_\pm =0,
\end{equation}
written here in a standard non-dimensional form. The parameter
$\delta$ measures the difference between the intra-species nonlinear
interaction constants: $\alpha_1+\delta$ corresponds to the
interaction among $\psi_+$ particles and $\alpha_1-\delta$ to the one
among $\psi_-$ particles. It is supposed that the two species are
labeled in such a way that $\delta>0$. $\alpha_2$ is the inter-species
interaction constant.  In the case of atomic condensates, the two
component order parameter may describe a two-species BEC such as
realized by considering for instance $^{87}$Rb in two hyperfine states
\cite{Mya97}, or a mixture of two elements \cite{Mod02}, or different
isotopes of the same atom \cite{Pap08}. In the case of a polariton
condensate, the components $\psi_+$ and $\psi_-$ account for the
pseudo-spin of the polariton which consists in a resonant
admixture of excitons and photons with spin projection $\pm 1$.

It is convenient to introduce spinor variables \cite{ktu-2005}
\begin{equation}\label{eq2}
    \left(
            \begin{array}{c}
              \psi_+ \\
              \psi_- \\
            \end{array}
          \right)=
          \sqrt{\rho}\, e^{i\Phi/2}\chi=\sqrt{\rho}\, e^{i\Phi/2}
          \left(
            \begin{array}{c}
              \cos\frac{\theta}2\,e^{-i\phi/2} \\
              \sin\frac{\theta}2\,e^{i\phi/2}  \\
            \end{array}
          \right).
\end{equation}
Here $\rho(x,t)=|\psi_+|^2+|\psi_-|^2$ denotes the total density of the
condensate and $\Phi(x,t)$ has the meaning of the velocity potential of its
in-phase motion; the angle $\theta(x,t)$ is the variable describing the
relative density of the two components
($\cos\theta=(|\psi_+|^2-|\psi_-|^2)/\rho$) and $\phi(x,t)$ is the
potential of their relative (counter-phase) motion. Accordingly, the
densities of the components of the condensate are given by
\begin{equation}\label{eq2c}
    \rho_+(x,t)=\rho\cos^2(\theta/2),\quad \rho_-(x,t)=\rho\sin^2(\theta/2),
\end{equation}
their phases are defined as
\begin{equation}\label{eq2a}
    \varphi_+(x,t)=\tfrac12(\Phi-\phi),\quad \varphi_-(x,t)=\tfrac12(\Phi+\phi),
\end{equation}
and the corresponding velocities of each components are
\begin{equation}\label{eq2d}
    v_+(x,t)=\nabla\varphi_+,\quad v_-(x,t)=\nabla\varphi_-.
\end{equation}
It is also convenient to introduce a unit vector $\mathbf{S}$
representing the spinor $\chi$:
\begin{equation}\label{eq2b}
    \mathbf{S}(x,t)=\chi^\dag \boldsymbol{\sigma}\chi=\left(
            \begin{array}{c}
              \sin{\theta}\cos\phi \\
              \sin\theta\sin\phi  \\
              \cos\theta
            \end{array}
          \right),
\end{equation}
where $\boldsymbol{\sigma}=(\sigma_x,\,\sigma_y,\,\sigma_z)^T$ is a
vector of Pauli matrices.  The vector $\mathbf{S}$ can be called the {\it
  polarization} vector of the two-component condensate.

Substitution of Eq.~(\ref{eq2}) into Eq.~(\ref{eq1}) yields the system
\begin{equation}\label{eq3}
    \begin{split}
    \rho_t&+\tfrac12[\rho(U-v\cos\theta)]_x=0,\\
    \Phi_t&-\frac{\cot\theta}{2\rho}(\rho\theta_x)_x
+\frac{\rho_x^2}{4\rho^2}-\frac{\rho_{xx}}{2\rho}+
    \tfrac14(\Phi_x^2+\theta_x^2+\phi_x^2)\\
&+\rho(\alpha_1+\alpha_2+\delta\cos\theta)=0,\\
    \rho\theta_t&+\tfrac12[(\rho v\sin\theta)_x+\rho U\theta_x]=0,\\
    \phi_t&-\frac1{2\rho\sin\theta}(\rho\theta_x)_x
+\tfrac12Uv-\rho[\delta+(\alpha_1-\alpha_2)\cos\theta]=0,
    \end{split}
\end{equation}
where $U=\Phi_x$ and $v=\phi_x$ are the mean and the relative
velocities: hence the velocities of the two components
are equal to $v_{\pm}=(U\pm v)/2$). In what follows, we shall consider
solitons and other nonlinear excitations corresponding to small
deviations from a uniform quiescent condensate and, to simplify the
treatment, we shall suppose that the two components in a non-excited
condensate have equal densities. This can be realized by two choices
$\theta\to\frac{\pi}2$ or $\theta\to\frac{3\pi}2$ for
$|x|\to\infty$. To be definite, we shall consider the case
$\theta\to\frac{\pi}2$ since another choice leads to similar results
when only signs of some parameters change.  The other parameters of
the wave must satisfy the boundary conditions
\begin{equation}\label{eq4}
    \rho\to\rho_0,\quad  U\to0,\quad v\to0 \quad\text{for}\quad |x|\to\infty.
\end{equation}
In the uniform state, the dependence of the phases on time is
described by the factors $\psi_{\pm}\propto\exp(-i\mu_{\pm}t)$, where
the chemical potentials $\mu_{\pm}$ are given by
\begin{equation}\label{eq4b}
\begin{split}
    &\mu_+=\tfrac18(U-v)^2+
\rho_0[(\alpha_1-\alpha_2+\delta)\cos^2(\theta_0/2)+\alpha_2],\\
    &\mu_-=\tfrac18(U+v)^2+
\rho_0[(\alpha_1-\alpha_2-\delta)\sin^2(\theta_0/2)+\alpha_2]
\end{split}
\end{equation}
(for future convenience we keep here the general notation for $\theta_0$).

First, we shall consider linear waves propagating along uniform
background. Linearization of the system (\ref{eq3}) with respect to
small variables $\rho'=\rho-\rho_0$, $\theta'=\theta-\pi/2$, $U$ and
$v$ yields
\begin{equation}\label{eq5}
    \begin{split}
    &\rho_t'+\tfrac12\rho_0U_x=0,\\
&U_t+(\alpha_1+\alpha_2)\rho'_x-\frac1{2\rho_0}\rho'_{xxx}-
\delta\rho_0\theta'_x=0,\\
    &\theta_t+\tfrac12v_x=0,\\
&v_t+\rho_0(\alpha_1-\alpha_2)\theta'_x-\tfrac12\theta'_{xxx}-\delta\rho'_x=0.
    \end{split}
\end{equation}
One can notice that for $\delta=0$ the system (\ref{eq5}) splits into
two pairs of independent equations: the first pair describes the
density waves with oscillations of $\rho$ and $U$ and the second pair
describes the polarization waves with oscillations of relative density
measured by $\theta$ and relative velocity $v$.  For $\delta\neq0$
these two modes are mixed but for convenience we will call them
``density'' and ``polarization'' waves if they transform into these
waves in the limit $\delta\to0$. If we look for the solution of the
system (\ref{eq5}) in the form of plane waves with all variables
proportional to $\exp[i(kx-\omega t)]$, then we readily get the
dispersion laws
\begin{equation}\label{eq6}
\begin{split}
    &\omega_d^2(k)=\tfrac12\rho_0\left(\alpha_1+
\sqrt{\alpha_2^2+\delta^2}\right)k^2+\tfrac14k^4,\\
    &\omega_p^2(k)=\tfrac12\rho_0\left(\alpha_1-
\sqrt{\alpha_2^2+\delta^2}\right)k^2+\tfrac14k^4,
\end{split}
\end{equation}
for the density and polarization modes, correspondingly. In the long
wavelength limit $k\to0$ we obtain the expressions for the density and
polarization sound velocities $c^2_{d,p}=\lim_{k\to 0}(\omega^2_{d,p}(k)/k^2)$
\begin{equation}\label{eq7}
\begin{split}
    &c_d^2=\tfrac12\rho_0\left(\alpha_1+\sqrt{\alpha_2^2+\delta^2}\right),\\
    &c_p^2=\tfrac12\rho_0\left(\alpha_1-\sqrt{\alpha_2^2+\delta^2}\right).
    \end{split}
\end{equation}
In the case where $\delta\ne 0$, the degree of mixture between density
and polarization waves can be evaluated by studying the dynamic
structure factor $S(k,\omega)$ of the system. At zero temperature
$S(k,\omega)=-\frac{1}{\pi} \Theta(\omega) \, \mbox{Im}\,
\chi(k,\omega)$ where $\Theta$ is the Heaviside step function and
$\chi(k,\omega)$ is the density response function (see, e.g.,
Refs.~\cite{Pin66} and \cite{Pit03}). $\chi(k,\omega)$\
characterizes how the density of the system responds to a weak
external scalar potential with wave vector $k$ and frequency
$\omega$. In presence of such a perturbation, using a trivial
modification of Eqs.~\eqref{eq5} accounting for the effect of the
external potential, one obtains
\begin{equation}\label{eq7a}
\chi(k,\omega)=\frac{Z_d(k)}{(\omega+i0^+)^2-\omega_d^2(k)} +
\frac{Z_p(k)}{(\omega+i0^+)^2-\omega_p^2(k)},
\end{equation}
with
\begin{equation}\label{eq7b}
\begin{split}
&Z_d(k)=\frac{\rho_0\, k^2}{2}
\left(1+\frac{\alpha_2}{\sqrt{\alpha_2^2+\delta^2}}\right)
\quad\mbox{and}\\
&Z_p(k)=\frac{\rho_0\, k^2}{2}
\left(1-\frac{\alpha_2}{\sqrt{\alpha_2^2+\delta^2}}\right).
\end{split}
\end{equation}
This yields
\begin{equation}\label{eq7c}
S(k,\omega)=
\frac{Z_d(k)}{2\,\omega_d(k)}\,\delta\Big(\omega-\omega_d(k)\Big) +
\frac{Z_p(k)}{2\,\omega_p(k)}\,\delta\Big(\omega-\omega_p(k)\Big).
\end{equation}
One has $\int_{\mathbb{R}}\omega S(k,\omega)d\omega =
\frac{1}{2}(Z_d(k)+Z_p(k))=\frac{1}{2}\rho_0 k^2$ in agreement with
the $f$-sum rule \cite{Pin66,Pit03}. In the case where $\delta=0$,
$Z_p(k)$ vanishes and the sum rule is exhausted by the peak at
$\omega_d(k)$: as stated above, this means that when $\delta=0$ the density
fluctuations are completely described by the branch with dispersion
$\omega_d(k)$. A further verification is that, in this case, the
Feynman relation holds: $\omega_d(k)=k^2/(2 \, S_k)$ where
$S_k=\int_{\mathbb{R}}S(k,\omega)d\omega$ \cite{Fey54}. When
$\delta\ne 0$ the relative contribution to the density fluctuation of
each branch can be evaluated by computing in which proportion the two
peaks in \eqref{eq7c} contribute to $S_k$. The ratio of these two
contributions to $S_k$ is easily evaluated in the low and large $k$
limits. Provided one does not go in the Manakov regime described
below, one sees in each of these two limiting cases that, for small
$\delta$, the contribution to $S_k$ of the mode with
dispersion $\omega_p(k)$ is lower by a factor of order
$(\delta/\alpha_2)^2$ than the contribution of the mode with
dispersion $\omega_d(k)$. Hence in the case of interest in the present
work, where $\delta$ is small compared to $\alpha_2$, one can
legitimately denote the branch with dispersion $\omega_d(k)$ the
density modulation branch, and the one with dispersion $\omega_p(k)$
the polarization modulation branch.

In all the present work we suppose that
\begin{equation}\label{eq8}
\alpha_1^2>\alpha_2^2+\delta^2,
\end{equation}
which is the condition of modulation stability of the polarization
mode (see, e.g., Ref.~ \cite{Pet02}). Expressions for dispersion laws
of linear waves propagating along binary condensates with non-equal
background densities of two components were found in
\cite{Fil05,gladush-2009}.

As we can see from Eq.~\eqref{eq7}, in the Manakov regime where all
nonlinearity constants are equal (i.e., $\delta=0$ and
$\alpha_1=\alpha_2$) the polarization sound velocity vanishes. In this
case the linear dispersion relation $\omega_p\approx c_pk$ can no
longer be considered as correctly describing the dispersion relation
in the long wavelength limit and the dispersive effects cannot be
considered as small even when $k\to 0$.  In the opposite configuration
where the difference $\alpha_1 -(\alpha_2^2+\delta^2)^{1/2}$ is large
enough, then the regime of linear dispersion
becomes of great importance when the
characteristic value of the wave-vector $k$ satisfies the condition
\begin{equation}\label{eq9}
    k^2\ll c_{d,p}^2.
\end{equation}
In this case, for linear waves propagating along the positive-$x$ direction,
the dispersion laws \eqref{eq6} can be approximated by
\begin{equation}\label{eq10}
    \omega_{d,p}(k)\cong c_{d,p} \, k+\frac1{8c_{d,p}}\, k^3.
\end{equation}
Correspondingly, the wave amplitude, say, $\rho'(x,t)$, satisfies the
linear equation
\begin{equation}\label{eq11}
    \rho'_t+c_{d,p}\, \rho'_x-\frac1{8c_{d,p}}\rho'_{xxx}=0,
\end{equation}
where the last term describes a small dispersive correction to the
propagation of pulses with constant sound speeds.

If the amplitude $\rho'$ is small but finite and such that the last
term in Eq.~(\ref{eq11}) has the same order of magnitude as the
leading nonlinear correction (typically, $\sim(\rho')^2$), then
nonlinear effects cannot be omitted for correctly
describing the propagation of the pulse. This issue is addressed in
the next section.

\section{Evolution equations for weakly nonlinear waves in
  a two-component BEC}

\subsection{Nonlinear density waves}\label{nldensity}

We now take into account
nonlinear effects in the propagation of a density pulse
with an accuracy up to second
order in the field variables $\rho',\,U,\,\theta',\,v$.  To simplify
the presentation, we omit the dispersive effects in a first
stage. Then the system (\ref{eq3}) reduces to
\begin{equation}\label{eq12}
    \begin{split}
    \rho'_t&+\tfrac12\rho_0U_x+\tfrac12\rho_0(\theta'v)_x+\tfrac12(\rho'U)_x=0,\\
    U_t&+(\alpha_1+\alpha_2)\rho'_x+\tfrac12(UU_x+vv_x)\\
    &-\delta\rho_0\theta'_x-\delta(\rho'\theta')_x=0,\\
    \theta'_t&+\tfrac12v_x+\frac1{2\rho_0}v\rho'_x+\tfrac12U\theta'_x=0,\\
    v_t&+\rho_0(\alpha_1-\alpha_2)\theta'_x+
\tfrac12(Uv)_x\\
&+(\alpha_1-\alpha_2)(\rho'\theta')_x- \delta\rho'_x=0.
    \end{split}
\end{equation}
Considering nonlinear density wave propagating in
the positive-$x$ direction in the standard
perturbation theory (see, e.g., \cite{kamch-2000}), one introduces the
stretched variables
\begin{equation}\label{eq13}
 \xi=\epsilon^{1/2}(x-c_dt),\quad \tau=\epsilon^{3/2}t
\end{equation}
and expand the fields variables $\rho',\,U,\,\theta',\,v$ in powers of
$\epsilon$:
\begin{equation}\label{eq14}
\begin{split}
& \rho'=\epsilon\rho^{(1)}+\epsilon^2\rho^{(2)}+\ldots,\quad
U=\epsilon U^{(1)}+\epsilon^2U^{(2)}+\ldots,\\
&\theta'=\epsilon\theta^{(1)}+\epsilon^2\theta^{(2)}+\ldots,\quad
v=\epsilon v^{(1)}+\epsilon^2v^{(2)}+\ldots.
\end{split}
\end{equation}
The corresponding expansion of Eqs.~(\ref{eq12}) yields, at leading
order, the consistent system of equations
\begin{equation}\label{eq15}
 \begin{split}
& -c_d\rho^{(1)}_{\xi}+\tfrac12\rho_0U^{(1)}_{\xi}=0,\\
& -c_dU^{(1)}_{\xi}+(\alpha_1+\alpha_2)\rho^{(1)}_{\xi}-
\delta\rho_0\theta^{(1)}_{\xi}=0,\\
& -c_d\theta^{(1)}_{\xi}+\tfrac12v^{(1)}_{\xi}=0,\\
& -c_dv^{(1)}_{\xi}+\rho_0(\alpha_1-\alpha_2)\theta^{(1)}_{\xi}-
\delta\rho^{(1)}_{\xi}=0.
\end{split}
\end{equation}
The determinant of this linear system is zero, and all variables can thus be
expressed it terms of one of them, $\rho^{(1)}$ for instance:
\begin{equation}\label{eq16}
\begin{split}
 &U^{(1)}=\frac{2c_d}{\rho_0}\rho^{(1)},\quad
\theta^{(1)}=\frac{\alpha_2-\sqrt{\alpha_2^2+\delta^2}}{\delta\rho_0}\rho^{(1)},
\\
&v^{(1)}=\frac{2c_d}{\delta\rho_0}\left(\alpha_2-
\sqrt{\alpha_2^2+\delta^2}\right)\rho^{(1)}.
\end{split}
\end{equation}
These are the relations actually realized in a linear density wave.
At next order in $\epsilon$ we get
\begin{equation}\label{eq17}
 \begin{split}
-c_d\rho^{(2)}_{\xi}&+\tfrac12\rho_0U^{(2)}_{\xi}\\
 &=-\rho^{(1)}_{\tau} -\tfrac12\rho_0(\theta^{(1)}v^{(1)})_{\xi}-\tfrac12(\rho^{(1)}U^{(1)})_{\xi},\\
-c_dU^{(2)}_{\xi}&+(\alpha_1+\alpha_2)\rho^{(2)}_{\xi}- \delta\rho_0\theta^{(2)}_{\xi}\\
 &= -U^{(1)}_{\tau}-\tfrac12(U^{(1)}U^{(1)}_{\xi}+v^{(1)}v^{(1)}_{\xi})+\delta(\rho^{(1)}\theta^{(1)})_{\xi},\\
-c_d\theta^{(2)}_{\xi}&+\tfrac12v^{(2)}_{\xi}\\
 &= -\theta^{(1)}_{\tau}-\frac1{2\rho_0}v^{(1)}\rho^{(1)}_{\xi}-\tfrac12U^{(1)}\theta^{(1)}_{\xi},\\
-c_dv^{(2)}_{\xi}&+\rho_0(\alpha_1+\alpha_2)\theta^{(2)}_{\xi}-\delta\rho^{(2)}_{\xi}\\
 &=-v^{(1)}_{\tau}-\tfrac12(U^{(1)}v^{(1)}_{\xi}-(\alpha_1-\alpha_2)(\rho^{(1)}\theta^{(1)})_{\xi}.
\end{split}
\end{equation}
As already observed at order ${\cal O}(\epsilon)$ [compare with 
Eqs.~(\ref{eq15})], the left-hand side of the system \eqref{eq17} has a
vanishing determinant, hence the expressions in this side are linearly
dependent. Therefore the expressions in the right-hand side must also
be linearly dependent, with the same proportionality coefficients
[which are the same as the ones already involved at order ${\cal
  O}(\epsilon)$ and explicitly written in Eqs.~ (\ref{eq16})]. This
condition yields the evolution equation
\begin{equation}\label{eq18}
 \rho^{(1)}_{\tau}+\frac{3(2\sqrt{\alpha_2^2+\delta^2}-\alpha_2)c_d}
{2\rho_0\sqrt{\alpha_2^2+\delta^2}}\rho^{(1)}\rho^{(1)}_{\xi}=0.
\end{equation}
We can now return to the previous coordinates $x$ and $t$ and take
into account the dispersion effects by simply adding the dispersive
term that was taken into account in Eq.~(\ref{eq11}). This is legitimate
because the only other
possible quadratic and dispersive term $\sim \rho^{(1)}\rho^{(1)}_{\xi\xi}$
which could have be missed in our dispersionless
approximation is forbidden since it does not
have the same symmetry with respect to the transformation $x\to-x$ as the
other terms. In other words, the inclusion
of this term would result in different equations for left- and
right-propagating waves. Thus, we arrive at the equation
\begin{equation}\label{eq19}
 \rho'_t+c_d\rho'_x+\frac{3(2\sqrt{\alpha_2^2+\delta^2}-\alpha_2)c_d}
{2\rho_0\sqrt{\alpha_2^2+\delta^2}}\rho'\rho'_{x}-\frac1{8c_{d}}\rho'_{xxx}=0,
\end{equation}
where $\rho'(x,t)=\rho(x,t)-\rho_0$
and we work at a level of
approximation where $\rho'(x,t)=\epsilon\rho^{(1)}(x,t)$. If the
solution of Eq.~(\ref{eq19}) is found, then the other variables can be
obtained with the use of relations (\ref{eq16}) which we rewrite for
completeness with the final notations:
\begin{equation}\label{eq19a}
\begin{split}
& U(x,t)=\frac{2c_d}{\rho_0}\rho^{\prime}(x,t),\quad
\theta^{\prime}(x,t)=\frac{\alpha_2-
\sqrt{\alpha_2^2+\delta^2}}{\delta\rho_0}\rho^{\prime}(x,t),\\
&v(x,t)=\frac{2c_d}{\delta\rho_0}
\left(\alpha_2-\sqrt{\alpha_2^2+\delta^2}\right)\rho^{\prime}(x,t).
\end{split}
\end{equation}
Equation (\ref{eq19}) is the KdV equation for weakly nonlinear density waves.
In the limit $\delta\to0$  we get the equation
\begin{equation}\label{eq20}
 \rho'_t+c_d\rho'_x+\frac{3c_d}{2\rho_0}\rho'\rho'_{x}
-\frac1{8c_{d}}\rho'_{xxx}=0,\quad
c_d=\sqrt{\tfrac12\rho_0(\alpha_1+\alpha_2)}
\end{equation}
which in the case $\alpha_2=\alpha_1$ reduces to the KdV equation for
shallow Manakov solitons.

\subsection{Nonlinear polarization waves: quadratic
  nonlinearity}\label{qnlpolar}

The equation of propagation of nonlinear polarization waves with account of
quadratic nonlinearity can be obtained by the same method. We
introduce the stretched variables
\begin{equation}\label{eq21}
 \xi=\epsilon^{1/2}(x-c_pt),\quad \tau=\epsilon^{3/2}t
\end{equation}
and make use of the series expansions (\ref{eq14}). At first order we
obtain the system (\ref{eq15}) with $c_d$ replaced by $c_p$ which
yields
\begin{equation}\label{eq22}
\begin{split}
&U^{(1)}=
\frac{2c_p\, \delta}{\alpha_2+\sqrt{\alpha_2^2+\delta^2}}\, \theta^{(1)},
\\
&\rho^{(1)}=\frac{\delta\, \rho_0}{\alpha_2+\sqrt{\alpha_2^2+\delta^2}}
\, \theta^{(1)},\\
&v^{(1)}={2c_p}\, \theta^{(1)}.
\end{split}
\end{equation}
These expressions are equivalent
to Eqs.~\eqref{eq16}---with $c_d$ replaced by $c_p$---but it is
now more convenient to express all variables in terms of $\theta^{(1)}$.
Tedious calculations at second order lead to the equation
\begin{equation}\label{eq23}
 \theta^{(1)}_{\tau}+\frac{3c_p(2\delta^2+\alpha_2^2-
\alpha_2\sqrt{\alpha_2^2+\delta^2})}
{2\delta\sqrt{\alpha_2^2+\delta^2}}\theta^{(1)}\theta^{(1)}_{\xi}=0.
\end{equation}
Returning to the coordinates $x$ and $t$ and taking the
dispersive effects into account we arrive again at the KdV equation
\begin{equation}\label{eq24}
 \theta'_{t}+c_p\theta'_x+\frac{3c_p(2\delta^2+\alpha_2^2
-\alpha_2\sqrt{\alpha_2^2+\delta^2})}
{2\delta\sqrt{\alpha_2^2+\delta^2}}\theta'\theta'_{x}
-\frac1{8c_p}\theta'_{xxx}=0\,
\end{equation}
describing the dynamics of weakly dispersive and weakly nonlinear
polarization waves. The other variables are expressed in terms
of $\theta'$ as follows:
\begin{equation}\label{eq24a}
\begin{split}
&U(x,t)=\frac{2c_p\delta}{\alpha_2+\sqrt{\alpha_2^2+\delta^2}}\,
\theta^{\prime}(x,t),\\
&\rho^{\prime}(x,t)=\frac{\delta\rho_0}{\alpha_2+\sqrt{\alpha_2^2+\delta^2}}
\, \theta^{\prime}(x,t),\\
&v(x,t)={2c_p}\, \theta^{\prime}(x,t).
\end{split}
\end{equation}
Now, contrarily to the case of density waves exposed in Sec.~ \ref{nldensity},
the coefficient of the
nonlinear term in (\ref{eq24}) vanishes in the limit $\delta\to0$: in
the limit $\delta\ll\alpha_2$ we get
\begin{equation}\label{eq25}
 \theta'_{t}+c_p\theta'_x+\frac{9c_p\delta}
{4\alpha_2}\theta'\theta'_{x}-\frac1{8c_p}\theta'_{xxx}=0,\quad
c_p=\sqrt{\frac{\rho_0(\alpha_1-\alpha_2)}2}.
\end{equation}
This means that if $\theta'\sim\delta\ll1$, then the level of
accuracy accepted here is not sufficient: the cubic nonlinearity terms
$\sim(\theta')^3$ neglected in the present treatment
have the same order of magnitude as the
quadratic term in \eqref{eq25}.
Thus, in the limit of small $\delta$ we have to
consider the next order of approximation.

\subsection{Nonlinear polarization waves: cubic nonlinearity}

As advocated in Sec.~\ref{qnlpolar}, cubic nonlinearities becomes
important when $\delta\sim\theta'$ is small and their
contribution can therefore be calculated from the system (\ref{eq3}) with
$\delta=0$. A series expansion up to the third order in the small
variables $\rho',\,U,\,\theta',\,v$ and up to the first order in the
derivatives of these quantities (we again postpone the inclusion of
dispersive effect for simplifying the presentation) reads
\begin{equation}\label{eq26}
    \begin{split}
    \rho'_t&+\tfrac12\rho_0U_x+\tfrac12\rho_0(\theta'v)_x+
\tfrac12(\rho'U)_x+\tfrac12(\rho'\theta' v)_x=0,\\
    U_t&+(\alpha_1+\alpha_2)\rho'_x+\tfrac12(UU_x+vv_x)\\
    &+\frac1{2\rho_0}\rho'_x(\theta'_x)^2-\frac{(\rho'_x)^3}{2\rho_0^3}=0,\\
    \theta'_t&+\tfrac12v_x+\frac1{2\rho_0}v\rho'_x+\tfrac12U\theta'_x\\
&-\frac1{2\rho_0^2}v\rho'\rho'_x-\tfrac12v\theta'\theta'_x-\tfrac14(\theta')^2v_x=0,\\
    v_t&+\rho_0(\alpha_1-\alpha_2)\theta'_x+\tfrac12(Uv)_x
+(\alpha_1-\alpha_2)(\rho'\theta')_x\\
&+\frac1{2\rho_0^2}(\rho'_x)^2\theta'_x
-\frac{\rho_0}2(\alpha_1-\alpha_2)(\theta')^2\theta'_x=0.
    \end{split}
\end{equation}
We now introduce the stretched variables
\begin{equation}\label{eq27}
 \xi=\epsilon^{1/2}(x-c_pt),\quad \tau=\epsilon^{5/2}t,\quad
c_p=\sqrt{\frac{\rho_0(\alpha_1-\alpha_2)}2},
\end{equation}
and use the series expansions (\ref{eq14}) up to the third order
terms in $\epsilon$. At order ${\cal O}(\epsilon)$ we get
\begin{equation}\label{eq28}
 \rho^{(1)}=0,\quad U^{(1)}=0,\quad v^{(1)}=2c_p\, \theta^{(1)}.
\end{equation}
At next order we obtain the algebraic relations
\begin{equation}\label{eq29}
\begin{split}
&\rho^{(2)}=-\frac{3c_p^2}{2\alpha_2}(\theta^{(1)})^2,\quad
U^{(2)}=-\frac{c_p(3\alpha_1+\alpha_2)}{2\alpha_2}\, (\theta^{(1)})^2,\\
&v^{(2)}=2c_p\, \theta^{(2)}.
\end{split}
\end{equation}
Finally, at order ${\cal O}(\epsilon^3)$ it is enough to consider the equations
for $\theta'$ and $v$ only:
\begin{equation}\label{eq30}
 \begin{split}
 -c_p\theta^{(3)}_{\xi}&+\tfrac12v^{(3)}_{\xi}
 =-\theta^{(1)}_{\tau}-\frac1{2\rho_0}\rho^{(2)}_{\xi}\\
&-\tfrac12U^{(2)}\theta^{(1)}_{\xi}+\tfrac12v^{(1)}\theta^{(1)}\theta^{(1)}_{\xi}
+\tfrac14(\theta^{(1)})^2v^{(1)}_{\xi}\\
-c_pv^{(3)}_{\xi}&+\rho_0(\alpha_1-\alpha_2)\theta^{(3)}_{\xi}
=-v^{(1)}_{\tau}
-\tfrac12(U^{(2)}v^{(1)})_{\xi}\\
&-(\alpha_1-\alpha_2)(\theta^{(1)}\rho^{(2)})_{\xi}
+\frac{\rho_0}2(\alpha_1-\alpha_2)(\theta^{(1)})^2\theta^{(1)}_{\xi}.
 \end{split}
\end{equation}
Again the expressions in the left-hand side are linearly dependent and
the compatibility condition for this system yields the evolution
equation which, with account of Eqs.~(\ref{eq28}) and (\ref{eq29}), can be
written as
\begin{equation}\label{eq31}
 \theta^{(1)}_{\tau}+\frac{3c_p}8 \left(1-9\frac{\alpha_1}{\alpha_2}\right)
(\theta^{(1)})^2\theta^{(1)}_{\xi}=0.
\end{equation}
Returning to the ($x,t$)-variables, and re-introducing dispersive
effects according to the procedure exposed in Sec.~ \ref{nldensity},
we arrive at the modified KdV equation
\begin{equation}\label{eq32}
 \theta'_{t}+c_p\theta'_x-\frac{3c_p}{8\alpha_2} \left(9{\alpha_1}-
{\alpha_2}\right)
\theta^{\prime2}\theta'_{x}-\frac1{8c_p}\theta'_{xxx}=0.
\end{equation}
At last, if $\delta$ is small and $\theta'\sim\delta$, we also have to take
into account the quadratic nonlinearity of Eq.~(\ref{eq25}) and corrections
of order ${\cal O}(\delta^2)$ to the velocity of the polarization sound:
\begin{equation}\label{eq33}
\begin{split}
\theta'_{t}&+\left(c_p-\frac{\rho_0\delta^2}{8c_p\alpha_2}\right)
\theta'_x+\frac{9\, c_p\, \delta}{4\alpha_2}\theta'\theta'_{x}\\
&-\frac{3c_p}{8\alpha_2} \left(9{\alpha_1}-{\alpha_2}\right)
\theta^{\prime2}\theta'_{x}-\frac1{8c_p}\theta'_{xxx}=0.
\end{split}
\end{equation}
This is Gardner equation describing the evolution of nonlinear polarization
pulses in a two-component condensate in the limit where
the intra-species interaction constants are close.
Once its solution is found, then
the other variables are expressed in terms of $\theta'$ by the
formulas which follow from Eqs.~(\ref{eq22}), (\ref{eq28}), and
(\ref{eq29}):
\begin{equation}\label{eq34}
\begin{split}
&     \rho'(x,t)=\frac1{2\alpha_2}\left(\rho_0\, \delta\, \theta'(x,t) -
3c_p^2\, \theta^{\prime2}(x,t)\right),\\
&    U(x,t)=\frac{c_p}{\alpha_2}
\left(\delta\, \theta'(x,t)
-\tfrac12(3\alpha_1+\alpha_2)\, \theta^{\prime2}(x,t)\right),\\
&v(x,t)=2 c_p\, \theta'(x,t).
\end{split}
\end{equation}
It is worth noticing that although in derivating the Gardner
equation (\ref{eq33}) we assumed the boundary condition $\theta'\to0$
as $|x|\to\infty$, this equation remains valid for the description of
the evolution of waves with boundary conditions
$\theta'\to\theta_{1,2}$ as $x\to\pm\infty$ provided the values
$\theta_{1,2}$ are small enough ($|\theta_{1,2}|\sim\delta\ll1$). As
we shall see, this additional freedom makes it possible to obtain
new types of solutions of the vector GP equation.

\section{Weakly nonlinear waves in a two-component BEC}

The system (\ref{eq1}) admits a number of solutions with different
properties depending on signs and values of
$\alpha_1,\alpha_2,\delta$. The possible solutions can also depend on
the background distributions of the condensate densities and
velocities. Our boundary conditions $\rho_{\pm}\to\rho_0/2$ exclude
such solutions as dark-bright solitons with vanishing at
$|x|\to\infty$ density of the bright soliton component. These
solutions have already been studied in the literature (see, e.g.,
Ref.~\cite{sp-1997,ka-2003}) and have been observed in experiments
\cite{anderson-2001,becker-2008,hamner-2011}.  However, nonlinear
coherent structures such as dark-dark solitons exist also in
situations with non-vanishing at infinity background densities (see,
e.g., \cite{hoefer-2011}) and we shall provide here several new
structures belonging to this class of solutions. To illustrate our
approach by a simple example, we shall start with well-known dark-dark
density solitons described in the limit of shallow solitons by the KdV
equation (\ref{eq19}).

\subsection{Density KdV solitons}

Evolution of density waves is described by the KdV equation
(\ref{eq19}) and its well-known soliton solution is given in this case
by the formula
\begin{equation}\label{eq35}
\begin{split}
    \rho'(x,t)&=-
\frac{2\rho_0\sqrt{\alpha_2^2+\delta^2}\,(c_d-V_s)}
{c_d(2\sqrt{\alpha_2^2+\delta^2}-\alpha_2)}\\
    &\times\frac1{\cosh^2\left[\sqrt{2c_d(c_d-V_s)}\,(x-V_st-x_0)\right]},
    \end{split}
\end{equation}
where $c_d$ is defined by the first of Eqs.~(\ref{eq7}).  From
Eqs.~(\ref{eq19a}) we can see that at $x\to\pm\infty$ the polarization
vector $\mathbf{S}$ lies in the $(S_x,\,S_y)$-plane, it rotates in
``southern hemisphere'' of the $\mathbf{S}$-space with changing $x$,
and its total rotation angle in the $(S_x,\,S_y)$-plane is equal to
\begin{equation}\label{eq36}
    \Delta\phi=\frac{4\sqrt{\alpha_2^2+\delta^2}(\sqrt{\alpha_2^2+\delta^2}
-\alpha_2)}{\delta(2\sqrt{\alpha_2^2+\delta^2}-\alpha_2)}
    \sqrt{1-\frac{V_s}{c_d}}.
\end{equation}
It goes to zero as $\Delta\phi\propto\sqrt{1-{V_s}/{c_d}}$ in the
limit $V_s\to c_d-0$ and as $\Delta\phi\propto\delta$ in the limit
$\delta\to0$.

If $\delta=0$, then the solution (\ref{eq35}) reduces to the limit of
shallow dark-dark soliton solution of the vector GP equation found in
\cite{gladush-2009}. At last, if in addition $\alpha_1=\alpha_2$, then
we reproduce the limit of shallow Manakov dark soliton
\cite{manakov}. It is important, that for $\delta=0$ the polarization
variable $\theta$ remains constant and in this case $v\equiv0$, so
that, hence, the polarization vector $\mathbf{S}$ does not vary.  We
shall call this solution the {\it density soliton} even for
$\delta\neq0$ when the polarization vector rotates according to
Eq.~(\ref{eq36}) since in the limit $\delta\to0$ it transforms to a
pure density mode.

\subsection{Polarization KdV solitons}

Now we turn to the polarization KdV solitons whose evolution is
described by Eq.~(\ref{eq24}). The soliton solution is given by
\begin{equation}\label{eq37}
\begin{split}
    \theta'(x,t)&=-\frac{2\delta\sqrt{\alpha_2^2+\delta^2}}
{2\delta^2+\alpha_2^2-\alpha_2\sqrt{\alpha_2^2+\delta^2}}\\
   & \times\frac{1-V_s/c_p}{\cosh^2\left[\sqrt{2c_p(c_p-V_s)}\,
(x-V_st-x_0)\right]},
\end{split}
\end{equation}
where $c_p$ is defined by the second of equations
(\ref{eq7}). Substitution of this expression into (\ref{eq24a}) allows
one to find other parameters of solution. Again, if the phase $\phi$ is
defined in such a way that $\phi=0$ at the center of the soliton, then
the polarization vector rotates, as we go along the soliton solution,
by the angle
\begin{equation}\label{eq38}
    \Delta\phi=-\frac{2\sqrt{2}\delta\sqrt{\alpha_2^2+\delta^2}}
{\delta(2\delta^2+\alpha_2^2-\alpha_2\sqrt{\alpha_2^2+\delta^2})}
    \sqrt{1-\frac{V_s}{c_d}}.
\end{equation}
It goes to zero as $\Delta\phi\propto\sqrt{1-{V_s}/{c_d}}$ in the
limit $V_s\to c_d-0$, however, for $\delta\to0$ this angle diverges as
$\Delta\phi\propto\delta^{-1}$ what shows that the polarization mode
ceases to exist in this limit. Such a behavior demonstrates drastic
difference between density and polarization modes. The same remark
refers to the amplitude of the polarization soliton (\ref{eq37}). We
can see that for small $\delta\ll|\alpha_2|$ the applicability
condition of the small amplitude approximation $|\theta'|\ll1$ is
satisfied provided that
\begin{equation}\label{eq39}
    1-\frac{V_s}{c_p}\ll\frac{\delta}{|\alpha_2|}.
\end{equation}
Even more heavy restriction follows from the condition that the
soliton amplitude must be much smaller than the coefficient in the
quadratic nonlinearity term so that we can neglect the cubic
nonlinearity terms:
\begin{equation}\label{eq40}
    1-\frac{V_s}{c_p}\ll\left(\frac{\delta}{\alpha_2}\right)^2.
\end{equation}
Thus, the applicability region of the KdV approximation is extremely
small for $\delta\ll|\alpha_2|$ and we must take into account the
cubic nonlinearity which corresponds to the Gardner equation
(\ref{eq33}).

\subsection{Polarization Gardner solitons}

The solution of the Gardner equation depends on signs and values of
the nonlinear interaction constants $\alpha_1,\,\alpha_2,\,\delta$.
In what follows we suppose that $\alpha_1>0$, $9\alpha_1-\alpha_2>0$,
$\alpha_2$ can be negative, and the condensate components are ordered
in such a way that $\delta>0$.

At first we assume that $\alpha_2>0$.  The soliton solution of the
Gardner equation (\ref{eq33}) is given by the formula (see, e.g.,
\cite{kamch-2012})
\begin{equation}\label{eq41}
    \theta'(x,t)=
\frac{\theta_1\theta_2}{\theta_1-(\theta_1-\theta_2)
\cosh^2[\sqrt{2c_pV}(x-V_st)]},
\end{equation}
where
\begin{equation}\label{eq42}
    c_p=\sqrt{\tfrac12\rho_0(\alpha_1-\alpha_2)},\quad
V_s=c_p-\frac{\rho_0\delta^2}{8c_p\alpha_2}-V,
\end{equation}
and
\begin{equation}\label{eq43}
    \theta_{1,2}=\frac{6\delta \pm2\sqrt{(3\delta )^2+
4\alpha_2 (9\alpha_1-\alpha_2)V/c_p}}{9\alpha_1-\alpha_2},
\end{equation}
where subscripts `1' and `2' correspond to the upper and lower signs,
respectively, $V$ is a free parameter defining the velocity and other
properties of the soliton solution (notice that $V$ measures the
soliton velocity in the reference frame moving with the sound velocity
equal to $c_p-{\rho_0\delta^2}/({8c_p\alpha_2})$). Substitution of
(\ref{eq41}) into Eqs.~(\ref{eq34}) yields the density and
polarization distributions of the polarization soliton.

Now the soliton's amplitude $\theta_1$ remains finite in the limit $\delta\to0$,
\begin{equation}\label{eq44}
    \left.\theta_1\right|_{\delta=0}=4
\sqrt{\frac{\alpha_2}{9\alpha_1-\alpha_2}\cdot\frac{V}{c_p}}
\end{equation}
and it is small for small enough $V$. The rotation angle of the
polarization vector along the soliton solution is given by the
expression
\begin{equation}\label{eq45}
    \Delta\phi=8\sqrt{\frac{2\alpha_2}{9\alpha_1-\alpha_2}}
\arctan\left(\sqrt{-\frac{\theta_1}{\theta_2}}\right)
\end{equation}
and it is finite for all $\delta$ and $V$.

\begin{figure}[ht]
\includegraphics[width=8cm]{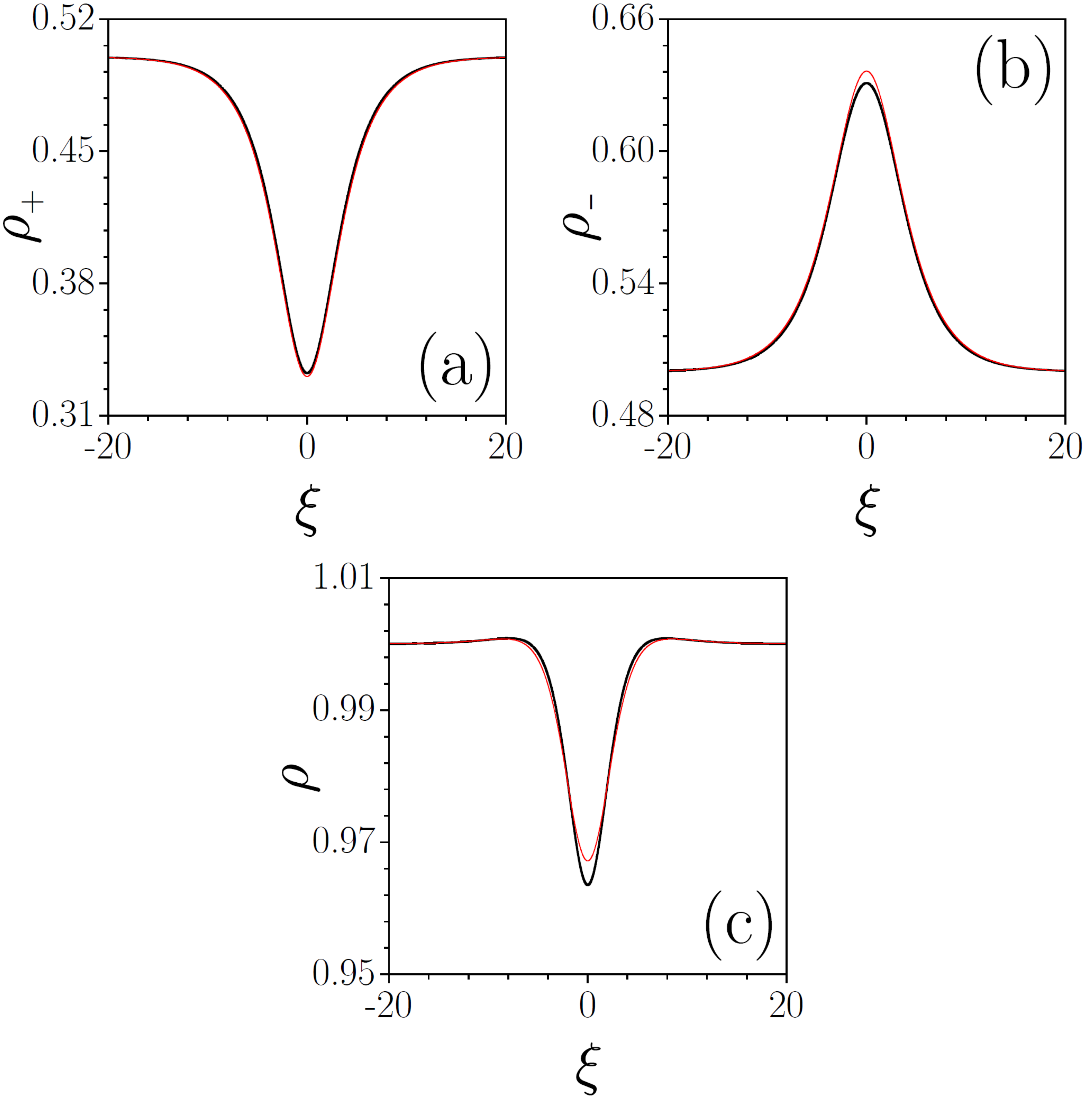}
\caption{(Color on line.) Dependence of (a) $\rho_+$, (b) $\rho_-$ and
  the total density (c) $\rho$ on $\xi=x-V_st$ for
  $\alpha_1=1.0,\,\alpha_2=0.6,\,\delta=0.05$ and velocity parameter
  $V=0.03$ (soliton velocity $V_s=0.416$). Analytical results in the
  Gardner approximation are shown by thick black lines and exact
  numerical results by red lines.}
\label{fig1}
\end{figure}
Densities of the two components of the soliton solution can be found
from Eqs.~(\ref{eq2c}).  Their plots are illustrated in Fig.~1
together with exact numerical solutions of the vector GP equations for
the same choice of parameters. Analogous plots for the flow velocities
of two components are shown in Fig.~2.
\begin{figure}[ht]
\includegraphics[width=8cm]{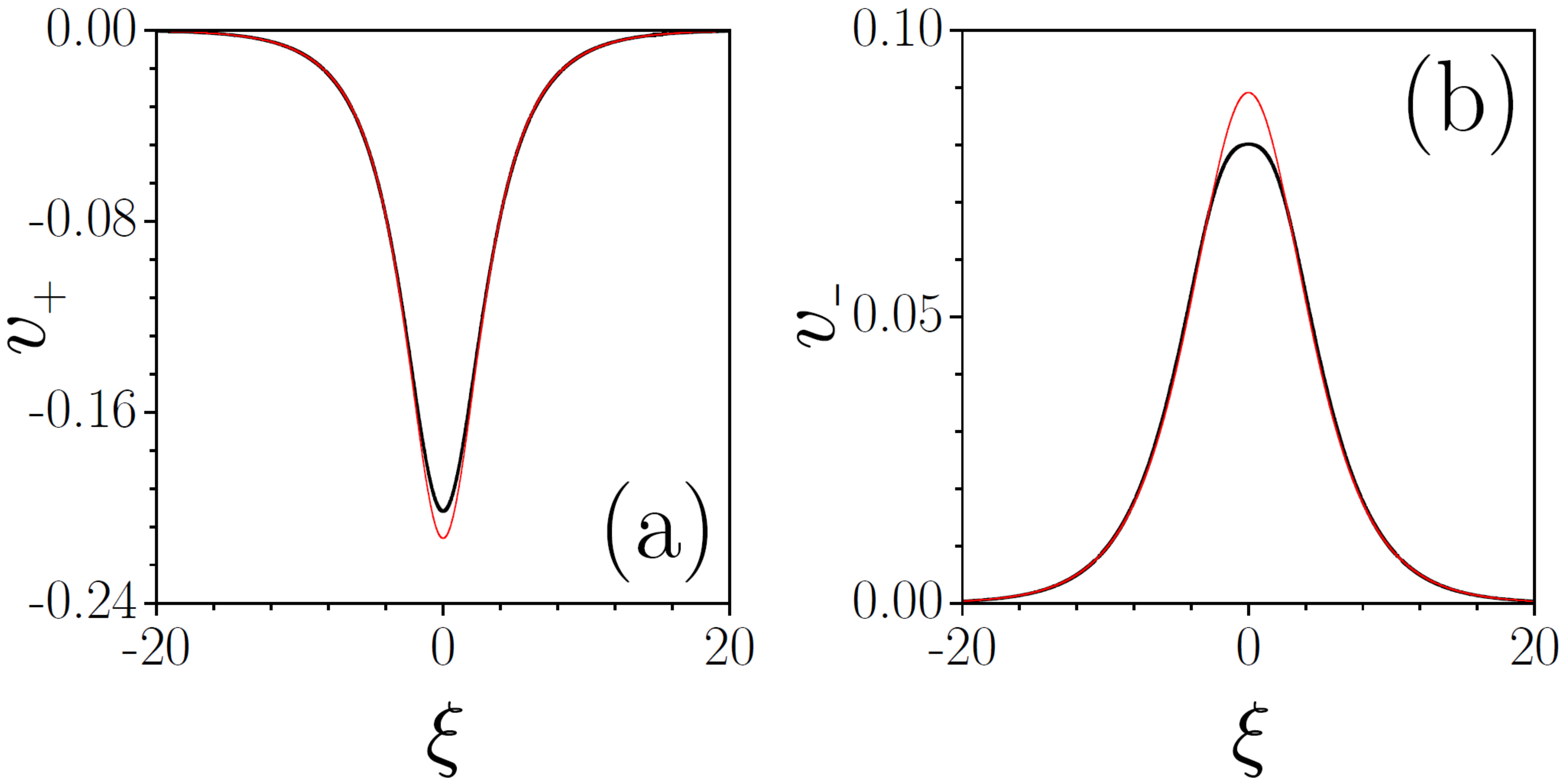}
\caption{(Color on line.) Dependence of (a) $v_+=(U-v)/2$ and (b)
  $v_-=(u+v)/2$ on $\xi=x-V_st$ for
  $\alpha_1=1.0,\,\alpha_2=0.6,\,\delta=0.05$ and velocity parameter
  $V=0.03$ (soliton velocity $V_s=0.416$). Analytical results in the
  Gardner approximation are shown by thick black lines and exact
  numerical results by red lines.}
\label{fig2}
\end{figure}

In the center of a soliton and for $V\ll
\rho_0\delta^2/(8\alpha_2c_p)$, that is for soliton's velocity $V_s$
close to the linear polarization wave velocity $c_p$, the condensate's
components densities are equal to
\begin{equation}\label{eq46}
    \begin{split}
    \rho_+=\frac{\rho_0}2
\Bigg\{1&-4\sqrt{\frac{\alpha_2}{9\alpha_1-\alpha_2}}
\cdot\sqrt{1-\frac{V_s}{c_p}}\\
&- \frac{12(\alpha_1-\alpha_2)}{9\alpha_1-\alpha_2}
\left(1-\frac{V}{c_p}\right)\Bigg\},\\
    \rho_-=\frac{\rho_0}2\Bigg\{1&+4\sqrt{\frac{\alpha_2}{9\alpha_1-\alpha_2}}
\cdot\sqrt{1-\frac{V_s}{c_p}}\\
&- \frac{12(\alpha_1-\alpha_2)}{9\alpha_1-\alpha_2}
\left(1-\frac{V}{c_p}\right)\Bigg\}.
    \end{split}
\end{equation}
These formulae agree with exact numerical solution of the vector GP
equations (\ref{eq1}) for $V_s$ close enough to $c_p$ (see Fig.~3).
\begin{figure}[ht]
\includegraphics[width=8cm]{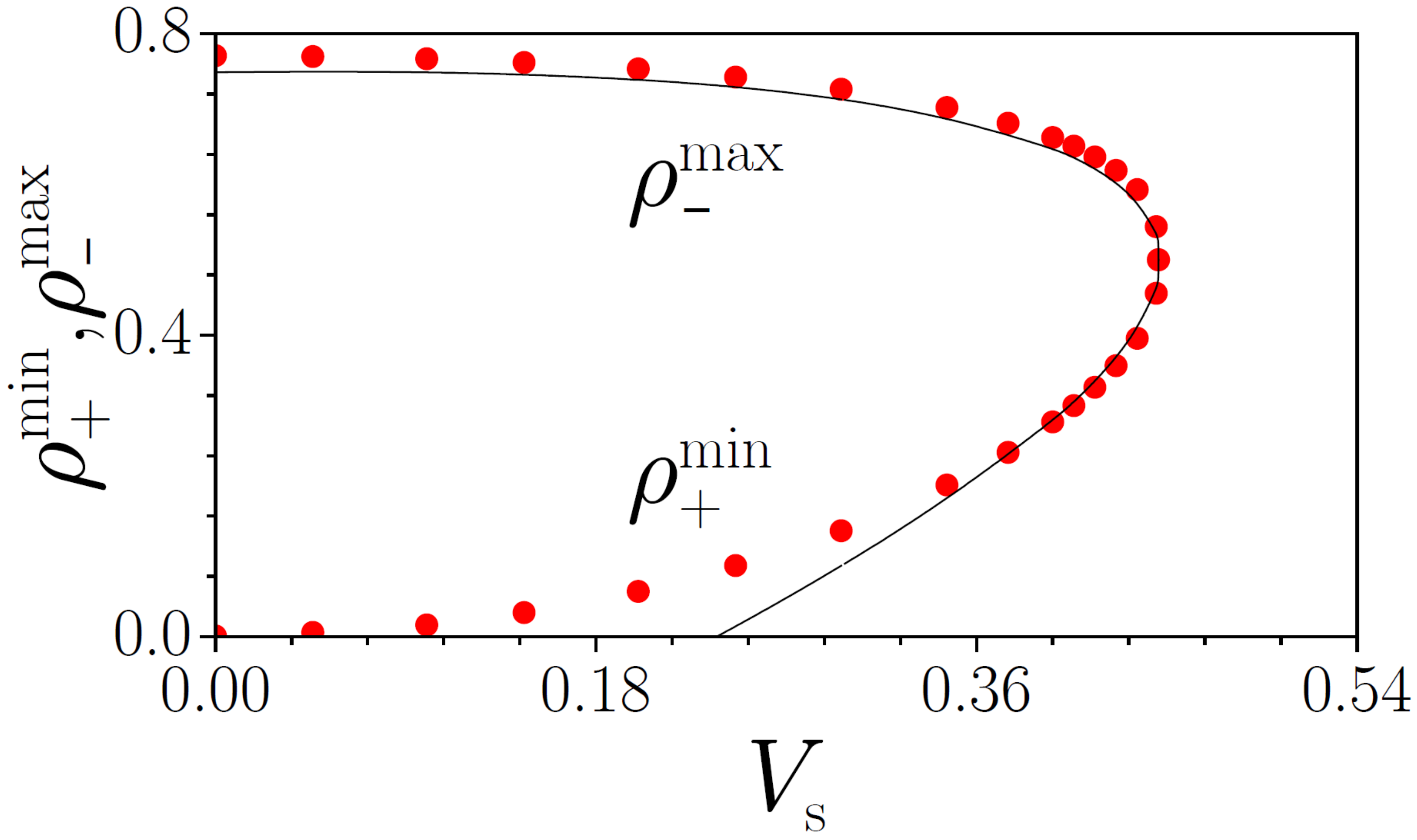}
\caption{(Color on line.) Dependence of two components densities at
  the center of soliton as functions of the soliton velocity $V_s$ for
  $\alpha_1=1.0,\,\alpha_2=0.6,\,\delta=0.05$. Analytical results in
  the Gardner approximation are shown by a solid line and exact
  numerical results by red dots.}
\label{fig3}
\end{figure}

If $\alpha_2<0$, the formula (\ref{eq41}) for the soliton solution
remains the same, but now in (\ref{eq43}) subscripts `1' and `2'
correspond to the lower and upper signs, respectively.

\subsection{Polarization algebraic solitons}

If $\alpha_2>0$, then there exists another type of soliton solutions
of the Gardner equation---so called {\it algebraic soliton} (see,
e.g., \cite{kamch-2012}). Such a solution is given by the expression
\begin{equation}\label{alg1}
    \theta'(x,t)=\theta_2+
\frac{\theta_1-\theta_2}{1+
\frac{c_p^2(9\alpha_1-\alpha_2)(\theta_1-\theta_2)^2}{8\alpha_2}
    (x-V_st-x_0)},
\end{equation}
where $\theta_2$ is a free non-zero background density parameter which
parameterizes the solution, and the other parameters are defined by
the equations
\begin{equation}\label{alg2}
    \theta_1=\frac{12\delta}{9\alpha_1-\alpha_2}-3\theta_2,
\end{equation}
\begin{equation}\label{alg3}
    V_s=c_p-\frac{\delta^2}{8c_p\alpha_2}+\frac{9c_p\delta}{8\alpha_2}-
\frac{3c_p(9\alpha_1-\alpha_2)}{8\alpha_2}\theta_2^2.
\end{equation}
In this case, for the background values of the angle
$\theta_2\sim\delta$, the soliton velocity departs from the sound
velocity to a value $\sim\delta$, too, and, hence, the amplitude of
the soliton is of the same order of magnitude. Thus, this is very
small amplitude and very wide soliton.

\subsection{Solibore solutions}

Besides soliton solutions, similar in many respects to the KdV
solitons, the Gardner equation has other types of solutions that are
absent in the KdV approximation. Here we shall consider one such a
solution, called ``solibore'' solution (see, e.g.,
\cite{holloway99,kamch-2012}). In some applications the solutions of
this kind are called ``kinks'' or ``half-solitons'' \cite{fsm-11}, however in hydrodynamics they
represent a degenerate case of a bore, i.e. of a solution that
connects, analogously to a shock wave, two flows with different
parameters.  Exactly this meaning for the condensate flows has the
solution that we are going to study here.

The solibore solution exists only for the case $\alpha_1>0$,\,
$\alpha_2<0$, if the background flow is modulationally stable.  In
this subsection we shall assume that these conditions are fulfilled.
In terms of the solution of the Gardner equation (\ref{eq33}) the
solibore solution relates the flows with different values of the angle
$\theta$ at left and right infinities $x\to\pm\infty$:
\begin{equation}\label{eq47}
    \theta'\to\theta_1\quad\text{at}\quad x\to+\infty\quad\text{and}\quad
\theta'\to\theta_2\quad\text{at}\quad x\to-\infty.
\end{equation}
This replaces the boundary condition $\theta\to\pi/2$ presumed
earlier, but the Gardner equation is still applicable provided
$|\theta_{1,2}-\pi/2|$ are small enough. In the solibore solution the
limiting values $\theta_{1,2}$ are related by the expression
\begin{equation}\label{eq48}
    \theta_1+\theta_2=\frac{3c_p\delta}{8|\alpha_2|}.
\end{equation}
For definiteness we suppose that $\theta_1>\theta_2$ and
$\theta'\to\theta_1$ at $x\to+\infty$; then the solibore solution of
the Gardner equation can be written as
\begin{equation}\label{eq49}
    \theta'(x,t)=\theta_1-\frac{\theta_1-\theta_2}
    {1+\exp\left[\sqrt{\frac{9\alpha_1-\alpha_2}{2|\alpha_2|}}
(\theta_1-\theta_2)(x-V_st-x_0)\right]},
\end{equation}
where the solibore velocity equals to
\begin{equation}\label{eq50}
\begin{split}
    V_s=c_p-\frac{\rho_0\delta^2}{8\alpha_2c_p}&+
\frac{(9\alpha_1-\alpha_2)c_p}{16\alpha_2}\\
   &\times \left[\left(\frac{6\delta}{9\alpha_1-\alpha_2}+\theta_1\right)^2
-3\theta_1^2\right].
\end{split}
\end{equation}
This solution is parameterized by the value of the angle $\theta_1$
at the right infinity $x\to+\infty$; then $\theta_2$ is defined by
Eq.~(\ref{eq48}). The other variables can be found by substitution of
Eq.~(\ref{eq49}) into Eqs.~(\ref{eq34}). We have illustrated the
solibore solution by plots in Figs.~4.  As one can see, this solution
represents a two-fluid flow which converts one component to another:
downstream the solibore (on the left side) the $\rho_+$-component
dominates and upstream the $\rho_-$-component dominates. In terms of
the total density, the solibore solution looks as an asymmetric dark
soliton moving with velocity $V_s$.  It is important to notice that
velocities of two components are different and do not vanish at
$x\to\pm\infty$.
\begin{figure}[ht]
\includegraphics[width=8cm]{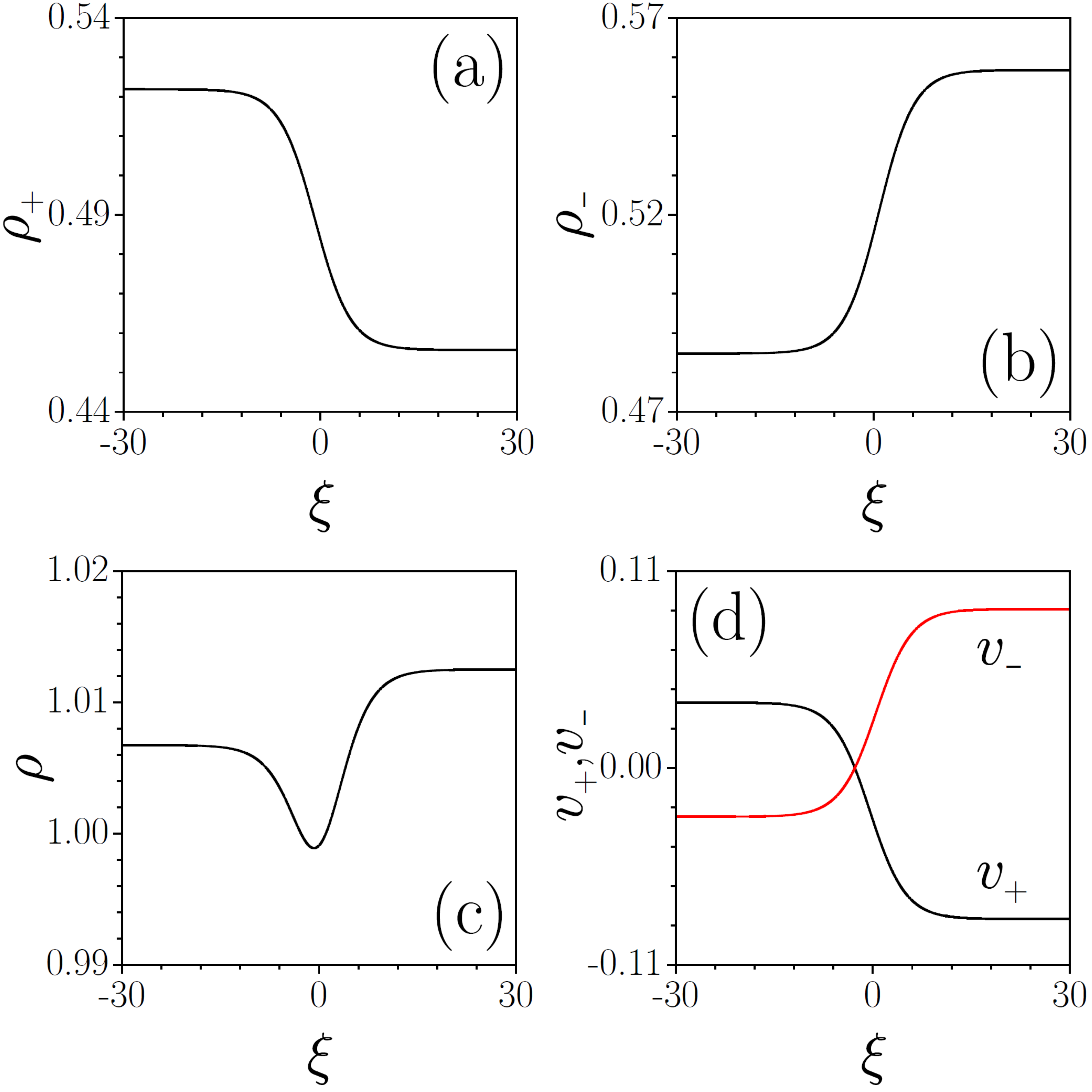}
\caption{(Color on line.) Dependence of $\rho_+$ (a), $\rho_-$ (b),
  the total density $\rho$ (c) and condensate's components
  velocities $v_+=(U-v)/2$, $v_-=(u+v)/2$ (d) on $\xi=x-V_st$ for
  $\alpha_1=1.0,\,\alpha_2=-0.6,\,\delta=0.05$ and $\theta_1=0.05$
  (solibore velocity $V_s=0.896$). Exact numerical solution of the
  vector GP equations cannot be distinguished from the analytical
  results obtained in the Gardner approximation and therefore they are
  not shown here.}
\label{fig4}
\end{figure}

\subsection{Breather solution}

As the last example of new wave structures supported by the vector GP
equation, we shall present the breather solution which appears as a
consequence of the corresponding solution of the Gardner equation,
found in \cite{pg-1997,gppt-2001}.  In our notation it exists if
$\alpha_2>0$ and can be written in the form
\begin{equation}\label{eq51}
\begin{split}
    \theta'(x,t)&=-\frac{2}{c_p}\sqrt{\frac{2\alpha_2}
{9\alpha_1-\alpha_2}}\\
&\times\frac{\prt}{\prt x}
    \arctan\frac{\kappa\cosh p\cos\Theta_b-k\cos q\sinh\Phi_b}
{\kappa\sinh p\sin\Theta_b+k\sin q\cosh\Phi_b},
\end{split}
\end{equation}
where $k$ and $\kappa$ are ``wavenumbers'' of an envelope and a
carrier wave, correspondingly,
\begin{equation}\label{eq52}
\begin{split}
   & k=\frac{3\delta c_p}{\sqrt{2\alpha_2(9\alpha_1-\alpha_2)}}
\frac{\sinh(2p)}{\cos^2q\cosh^2p+\sin^2q\sinh^2p},\\
   & \kappa=\frac{3\delta c_p}{\sqrt{2\alpha_2(9\alpha_1-\alpha_2)}}
\frac{\sin(2q)}{\cos^2q\cosh^2p+\sin^2q\sinh^2p},
\end{split}
\end{equation}
and velocities of the envelope and carrier wave are given by the expressions
\begin{equation}\label{eq53}
    V_b=c_p-\frac{\rho_0\delta^2}{8c_p\alpha_2}-\frac{3\kappa^2-k^2}{8c_p},\quad
    V_i=c_p-\frac{\rho_0\delta^2}{8c_p\alpha_2}-\frac{\kappa^2-3k^2}{8c_p},
\end{equation}
while the phases are defined as
\begin{equation}\label{eq54}
    \Theta_b=k(v-V_bt)+\Theta_0,\quad \Phi_b=\kappa(x-V_it)+\Phi_0.
\end{equation}
Substitution of (\ref{eq51}) into Eqs.~(\ref{eq2c}) yields the
densities of two components in the breather solution of vector GP
equation in the Gardner approximation. This solution is illustrated in
Fig.~5.  It describes a non-stationary propagation of a nonlinear wave
packet with envelope velocity $V_b$.
\begin{figure}[ht]
\includegraphics[width=8cm]{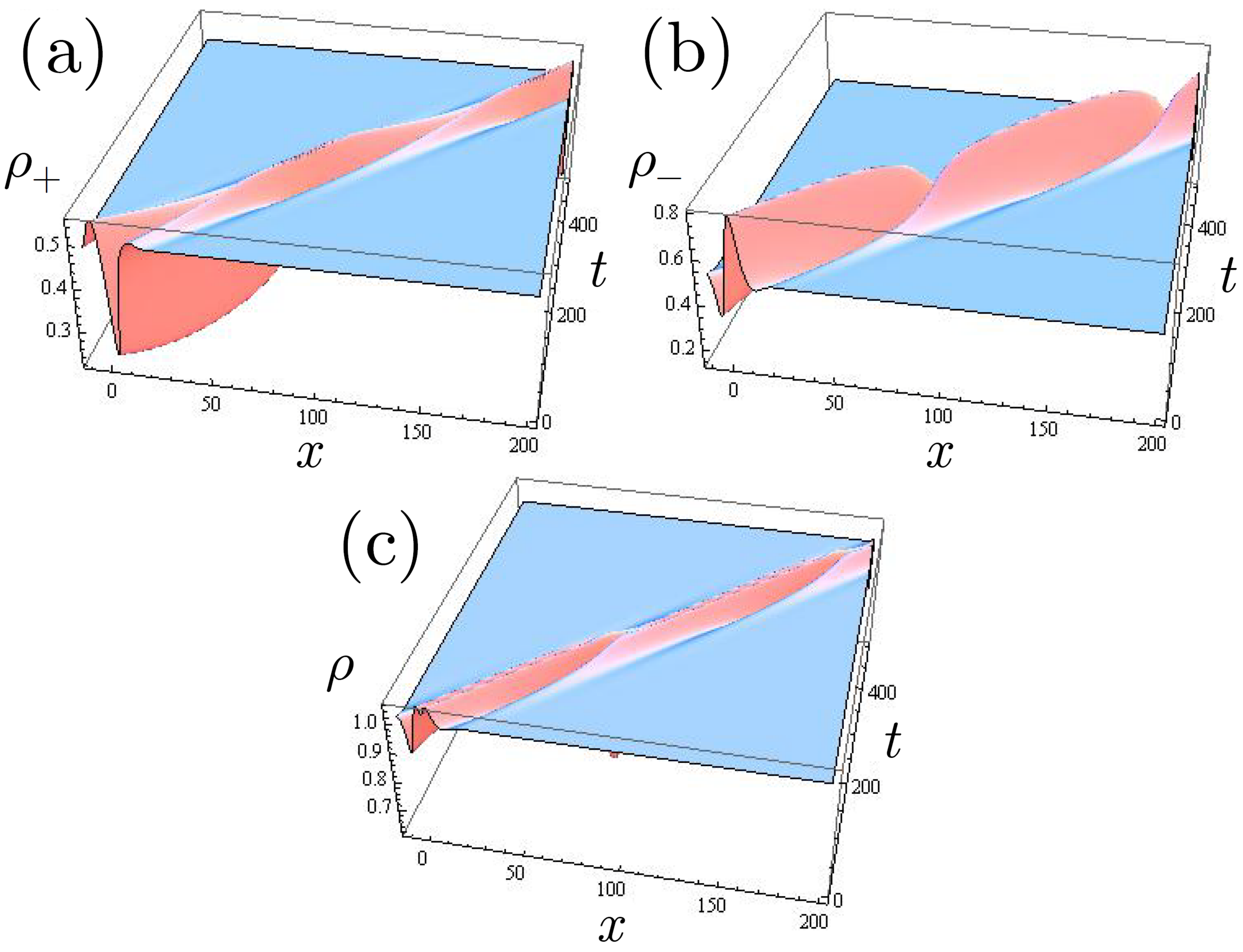}
\caption{(Color on line.) Plots of $\rho_+$ (a), $\rho_-$ (b), and the
  total density $\rho$ (c) as functions of $x$ and $t$ for
  $\alpha_1=1.0,\,\alpha_2=0.6,\,\delta=0.4$.  }
\label{fig5}
\end{figure}

\section{Conclusion}

In this paper, we have shown that for vector GP equation there exists
a region of values of the nonlinearity constants for which this quite
complicated system of equations can be reduced to a single nonlinear
evolution equation---the so-called Gardner equation. This equation
describes the evolution of the nonlinear polarization excitations,
that is of such excitations which in the linear limit reduce to the
polarization sound waves with constant total density of the
condensate. Such an approach leads to several new types of nonlinear
waves in the two-component condensates---algebraic solitons,
solibores, breathers.  Numerical solution of the vector GP equation
with corresponding initial data demonstrates that these new types of
nonlinear excitations propagate without change of their properties
during evolution, that is they are stable with respect to decay to
more elementary excitations. Although analytical description of the
wave pattern exists at the moment in the small amplitude approximation
only, similar large amplitude wave patterns can be described as exact
numerical solutions of the vector GP equation. We expect that these
new excitations can be generated in experiments by the flow of the
two-component condensate past obstacles or by the phase and density
engineering methods applied to the two-component condensate.

\begin{acknowledgments}
AMK thanks Laboratoire de Physique Th\'eorique et Mod\`eles
Statistiques, Universit\'e Paris-Sud, Orsay, where this work was
started, for kind hospitality.
This work was supported by the French ANR under grant
n$^\circ$ ANR-11-IDEX-0003-02 (Inter-Labex grant QEAGE).
\end{acknowledgments}

\end{document}